\documentclass[%
pre,
twocolumn,
times,
 amsmath,amssymb,
 aps,
]{revtex4-2}

\usepackage{lineno}%,hyperref}
\usepackage[colorlinks,linkcolor=black,citecolor=black,urlcolor=black]{hyperref}
\usepackage{graphicx}% Include figure files
\usepackage{dcolumn}% Align table columns on decimal point
\usepackage{bm}% bold math
\usepackage[utf8]{inputenc}
\usepackage{appendix}
\usepackage{todonotes}
\usepackage{soul}
\usepackage{xcolor}
\usepackage{lineno}%,hyperref}
\usepackage{comment}

\newcommand{\ri}{\mathrm{i}}

\newcommand{\diag}{\mathrm{diag}}

\begin{document}

%\preprint{}

\title{Rigorous Hydrodynamics from Linear Boltzmann Equations \\ and Viscosity-Capillarity Balance}

\author{Florian Kogelbauer}\thanks{Corresponding author}
 \email{flroiank@ethz.ch}
\author{Ilya Karlin}%\thanks{Corresponding author}
 \email{ikarlin@ethz.ch}
\affiliation{
 Department of Mechanical and Process Engineering, ETH Zurich,  CH-8092 Zurich, Switzerland 
}

\date{\today}

\begin{abstract}
Exact closure for hydrodynamic variables is rigorously derived from the linear Boltzmann kinetic equation. 
Our approach, based on spectral theory, structural properties of eigenvectors and the theory of slow manifolds, allows us to define a unique, optimal reduction in phase space close to equilibrium.
The hydrodynamically constrained system induces a modification of entropy that ensures pure dissipation on the hydrodynamic manifold, which is interpreted as a non-local variant of Korteweg's theory of viscosity-capillarity balance. The rigorous hydrodynamic equations are exemplified on the Knudsen minimum paradox in a channel flow.
%
%We propose a general mechanism to 
%rigorously derive an exact, closed system for the hydrodynamic variables from linear kinetic models. Our approach, based on spectral theory, structural properties of eigenvectors and the theory of slow manifolds, allows us to define a unique, optimal reduction in phase space close to the global equilibrium distribution.
%The hydrodynamically constrained system induces a modification of entropy that ensures pure viscous dissipation on the hydrodynamic manifold, which is interpreted as a non-local variant of Korteweg's theory of viscosity-capillarity balance. \textcolor{blue}{We exemplify the theory on the Knudsen minimum paradox.}
\end{abstract}

\maketitle

\section{Introduction}
\label{sec:intro}
The closure problem for kinetic equations constitutes one of the fundamental challenges in statistical physics. Based on the kinetic theory of Maxwell and Boltzmann, Hilbert \cite{hilbert1912grundzuge} asked a humble question: Is there a self-consistent way to derive the evolution of macroscopic variables directly from particle-based models, most prominently, the Boltzmann equation?

%\todo[inline, color = green ]{In fact, we expect that the exact summation of the Chapman--Enskog series is equivalent to the spectral closure.}

Classically, fluid models such as the Euler equations or the Navier--Stokes equations are derived from kinetic equations through expansions in Knudsen number - the famous Chapman--Enskog expansion \cite{chapman1990mathematical} - and were shown to be consistent with well-established local fluid models in the high-collision limit \cite{saint2009hydrodynamic}. Higher-order terms in Knudsen number, such as the Burnett equation, however, exhibit nonphysical behavior, such as instabilities \cite{bobylev1982chapman}, and cannot be extended to the medium- or low-collisional regime \cite{hadjiconstantinou2006limits}. Furthermore, there exist thermodynamic effects, such as thermally induced creep or Knudsen diffusion, which cannot be explained within the Navier--Stokes equations \cite{gorban2014hilbert}. Nevertheless, a reduced description in terms of finitely-many macroscopic quantities is highly desirable both in the light of moments closures as well as for efficient numerical methods \cite{levermore1996moment}.

To overcome these deficiencies, several modifications to existing fluid models where proposed, including adaptations of integro-differential operators \cite{PhysRevA.40.7193}, various discretization of the Boltzmann equation \cite{grad1949kinetic,shan2006kinetic}, modification based on general rheological considerations \cite{brenner2005navier} or, recently, models based on machine learning \cite{Candi2023,han2019uniformly}. We also mention the regularization of higher-order hydrodynamics obtained from the Chapman--Enskog expansion presented in \cite{bobylev2006instabilities}. A uniform, self-consistent mechanism that links any kinetic model to the dynamics of macroscopic variables, however, still lacks a solid foundation.

Guided by the principle of slow manifold reduction \cite{mckean1969simple,gorban1994method,gorban2014hilbert} we present a unified approach to deduce consistent hydrodynamic laws without any smallness assumption on the mean-free path or related quantities (e.g. Knudsen minimum), relying solely on the spectrum of the linear part in combination with structural properties of the associated eigenspaces. We will refer to this special, dynamically optimal closure as \textit{slow spectral closure}, which was first calculated for the BGK collision model in \cite{kogelbauerBGKspectral1,kogelbauerBGKspectral2}. While perturbative numerical investigations \cite{PhysRevLett.100.214503,colangeli2009boltzmann} have been carried out up to the level of Maxwell molecules, the alleged relation between a slow manifold approximation and a global fluid closure, is still poorly understood \cite{gorban2014hilbert}.

We emphasize that hydrodynamics obtained through the slow spectral closure are optimal in the following sense: The deviation of the hydrodynamically closed system from a general solution to the kinetic model is minimal over time and a general kinetic trajectory approaches the spectrally closed system exponentially fast in time. In particular, any other closure, be it one obtained from the truncated Chapman--Enskog expansion or by an ad-hoc closure procedure, will necessarily deviate more from a true solution than the spectrally closed hydrodynamics.

%\todo[inline, color = green]{@Ilya: could you please write a paragraph here linking the previous one to our findings (or indicate what I should write) plus if we should cite some more people here?}

In this work, we take advantage of a specific commutation property of the Boltzmann equation in frequency space in conjunction with spectral calculus to derive the algebraic form of the exact, invariant closure relations for the macroscopic variables. The resulting hydrodynamics are formulated in terms of generalized transport coefficients, they are inherently non-local in nature and provide a self-consistent set of equations for the density, the velocity field and the temperature. Based on the global dissipation relation that links the decay rate of a hydrodynamic mode to its wave number, we apply the resulting spectrally-closed dynamics to the question of viscosity-capillarity and hydrodynamic entropy.

The paper is organized as follows: 
In  Section \ref{sec:exact} we recall general spectral properties of linear Boltzmann-type kinetic operators and discuss a commutation property of the kinetic operator in frequency space. The action of a symmetry group implies the general structure of the hydrodynamic equations for the five macroscopic variables. 
Section III is concerned with implication of the hydrodynamic closure to viscosity-capillarity balance and the question of hydrodynamic entropy. We show that the spectral closure implies a wave-number dependent modification of entropy and dissipation which leads to an Onsager-type reciprocity relation.
In Section \ref{sec:knudsen}, we apply the new hydrodynamic equations to the Knudsen minimum problem. We recover the minimal flow in dependence of inverse Knudsen number to high accuracy and correctly predict the asymptotic behavior for zero Knudsen number.
Finally, some conclusions are drawn in Section \ref{sec:conclusion}.

\section{Exact hydrodynamics from the linearized Boltzmann equation}
\label{sec:exact}

\subsection{Invariant closure problem}
\label{eq:invariance}

We will be concerned with a general linear kinetic equation of the form,
\begin{equation}\label{maineq}
    {\partial_t f} = \mathcal{L}f,\quad \mathcal{L}=\mathcal{S}+\mathcal{Q},
\end{equation}
for an unknown distribution function $f$,  the free-flight operator $\mathcal{S} = - \bm{v}\cdot\nabla$ and the non-positive semi-definite linear collision operator $\mathcal{Q}=\mathcal{Q}^*$. Equation \eqref{maineq} is, e.g., obtained as the linearization of a Boltzmann-type equation around a global Maxwellian with respect to the inner product \begin{equation}
   \langle f,g \rangle = \int_{\mathbb{R}^3}d\mu(\bm{v})f(\bm{v})g^{*}(\bm{v}),
\end{equation}
where $d\mu = (2\pi)^{-\frac{3}{2}}e^{-\frac{{v}^2}{2}}d\bm{v}$ is the normalized Gaussian measure. We stress, however, that our considerations apply to any linear operator satisfying the conditions specified below.\\
We denote the projection onto the invariants of the collision operator as $\mathbb{P}$, which satisfies $\mathbb{P}\mathcal{Q}=\mathcal{Q}\mathbb{P}=0$, defining locally conversed quantities of \eqref{maineq} {and write  $\mathbb{P}^{\perp}=1-\mathbb{P}$}. While the reduction procedure presented below is not limited to a specific number of moments, we assume that there are five associated moments and denote them as
\begin{equation}
    {e} = \left(1,\bm{v},\frac{{v}^2-3}{\sqrt{6}}\right)
\end{equation}
such that $\mathbb{P}f = \langle f,{e}\rangle {e}$. The five macroscopic variables, the mass density, the velocity field and the temperature are then given as
\begin{equation}
{\left(\rho,\bm{u},\sqrt{\frac{3}{2}}T\right) = \langle f,{e}\rangle. }
\end{equation}

A self-consistent set of hydrodynamic equations derived from \eqref{maineq} is equivalent to the existence of a linear closure operator $\mathcal{C}$
%$\mathcal{C}:\text{range }\mathbb{P}\to \text{range }\mathbb{P}^{\perp}$, 
satisfying a quadratic restriction of being exact closure (the \emph{invariance equation}, see \cite{gorban2005invariant}),
\begin{equation}\label{eq:invariance}
	(\mathcal{C}\mathbb{P}-\mathbb{P}^{\perp})\mathcal{L}(1+\mathcal{C})= 0.
\end{equation}
%Higher-order moments are expressed a linear combination of the hydrodynamic moments. 
The subsequent analysis is based on the basic insight that an invariant manifold for the dynamics implies a closure operator for \eqref{maineq}. The eigenvectors associated to the hydrodynamic modes 
%(all having negative real part) 
span a slow subspace and any trajectory converges to this slow manifold exponentially fast \cite{Cab2003}, thus rendering the \textit{spectral closure} associated to the eigenvalues of $\mathcal{L}$ dynamically optimal.
%In the following, we detail the derivation of the spectral closure based on a commutation property and general spectral properties of $\mathcal{L}$.  

%\todo[inline, color=green]{Put invariance principle and slow manifold here, shorter, no details, outline of how we derive the H matrix, calculation steps}
%\todo[inline, color=yellow]{On another thought, let's leave it as is for the time being, the invariance principle story...}

\subsection{Spectral problem for linearized Boltzmann equation}
\label{sec:spectral}
Subsequent computations shall be carried out in Fourier space. Denoting the wave vector as $\bm{k}$ and writing $k=|\bm{k}|$ for the wave number, the linear evolution operator \eqref{maineq} in frequency space reads,
\begin{equation}\label{eq:Lk}
\mathcal{L}_{\bm{k}} = -\ri\bm{k}\cdot\bm{v} +\mathcal{Q},
\end{equation}
and we shall drop the index $\bm{k}$ to ease notation, wherever this does not lead to confusion.
We emphasize that, while the spectral structure of the linear collision operator $\mathcal{Q}$ is known explicitly up to the level of Maxwell molecules \cite{bobylev1988theory}, the spectral structure of the full wave-vector dependent operator $\mathcal{L}_{\bm{k}}$ is much more involved.

For any $\bm{O}$ orthogonal, define the unitary operator \begin{equation}
 { U_{\bm{O}}f(\bm{v}) = f(\bm{O}\bm{v}), \ U_{\bm{O}}^*=U_{\bm{O}^T},}
\end{equation}
and assume that the linear kinetic operator $\mathcal{L}_{\bm{k}}$ satisfies the commutation property,
\begin{equation}\label{commutation}
    U_{\bm{O}}\mathcal{L}_{\bm{k}} = \mathcal{L}_{\bm{O}^T\bm{k}}U_{\bm{O}}.
\end{equation}
Any linear Boltzmann collision operator of the form,
%\begin{small}
\begin{equation}\label{defLQ}
\begin{split}
    \mathcal{Q} = \int_{\mathbb{S}^2}\int_{\mathbb{R}^3}d\mu(\bm{v}_{*}) d\bm{\omega} B
    [f(\bm{v}_{*}')+f(\bm{v}')-f(\bm{v}_{*})-f(\bm{v})],
\end{split}
\end{equation}
%\end{small}
where $d\bm{\omega}$ is the element of the unit sphere, the pre- and post-collision velocities are $\bm{v}' =\bm{v}-(\bm{v}-\bm{v}_{*})\cdot\boldsymbol{\omega} \boldsymbol{\omega}$, $\bm{v}'_{*} =\bm{v}_{*}+(\bm{v}-\bm{v}_{*})\cdot \boldsymbol{\omega}\boldsymbol{\omega}$, 
and the collision kernel $B$ depends only on $|\bm{v}-\bm{v}_*|$ and the deflection angle  between $\bm{v}-\bm{v}_*$ and $\bm{v}'-\bm{v}_*'$,
satisfies condition \eqref{commutation}.  Indeed, the collision operator satisfies $[U_{\bm{O}},\mathcal{Q}]= 0$ by a rotational change of coordinates in the integral, while
{\begin{equation}
   -\ri\bm{k}\cdot \bm{v}f(\bm{O}\bm{v})  = (U_{\bm{O}}\mathcal{S}_{\bm{O}\bm{k}}f)(\bm{v}).
\end{equation}
}
For details, we refer to Appendix \ref{app:commprop}. The same holds true for the linear BGK operator \cite{bhatnagar1954model}, and, more generally, for any linear quasi-equilibrium approximation \cite{gorban1994general}. Properties similar to \eqref{commutation} are important tool in the spectral theory of the Boltzmann equation \cite{ellis1975first}.

\begin{figure}
    \centering
    \includegraphics[width=1\linewidth]{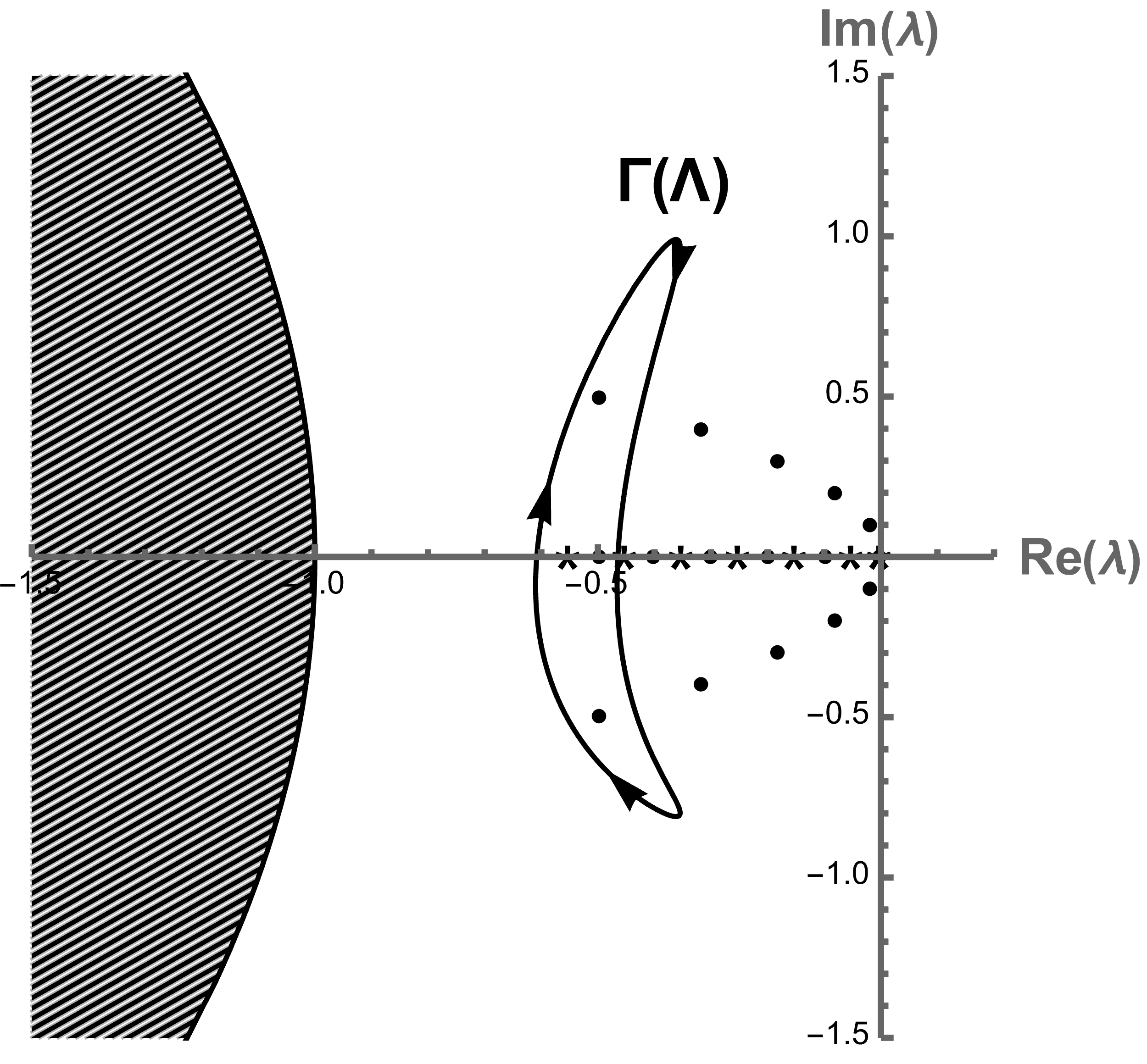}
    \caption{\footnotesize{Schematics of the spectrum (essential plus small wave number discrete) of the Boltzmann hard-spheres model. The essential spectrum (dashed area) extends all the way through the spectral plane, while for small $k$, five branches of hydrodynamic modes emerge from the origin. 
    The contour $\Gamma (\Lambda)$ encircles parts of the discrete spectrum to define the Riesz projection \eqref{eq:Riesz}.}}
   \label{spec1}
\end{figure}

We recall that, for each wave vector, the spectrum of $\mathcal{L}_{\bm{k}}$ is located entirely in the left half-pane, i.e., only has non-positive real parts, (with a five-fold degenerate eigenvalue at zero for $k=0$) and only depends on $k$. Classical spectral theory of the Boltzmann equation \cite{ellis1975first,cercignani1988boltzmann,dudynski2013} shows that the essential spectrum depends on the decay properties of the collision kernel and is independent on wave number, while - for a fixed wave number - the spectrum only consists entirely of isolated eigenvalues above the essential spectrum. Five branches of discrete eigenvalues (counted according to multiplicity) emerge from the origin, see Fig.\ \ref{spec1}:
\begin{equation}\label{LambdaHydro}
    \Lambda(k) = \{\lambda_{\rm s}(k), \lambda_{\rm s}(k), \lambda_{\rm d}(k),\lambda_{\rm a}(k),\lambda_{\rm a}^{*}(k)\},
\end{equation}
for the double degenerate, yet semi-simple, real shear mode $\lambda_{\rm s}$, the real diffusion mode $\lambda_{\rm d}$ and the pair of complex conjugated acoustic modes $\lambda_{\rm a},\lambda_{\rm a}^*$. As the wave number increases, the hydrodynamic modes move from the origin (center space of collision invariants) towards the essential spectrum, decreasing in real part with higher wave numbers. Each eigenvalue branch $\lambda_{j}(k)$ \eqref{LambdaHydro} exists 
%for a certain range of wave numbers 
up to a critical wave number, $k_{{\rm crit},j}>0$, at which the discrete spectrum merges into the essential spectrum
%$k\in[0,k_{{\rm crit},j}]$, 
\cite{dudynski2013}. The phenomenon of criticality in the context of kinetic dispersion relations was observed in \cite{karlin1997gradient,karlin2002grad,colangeli2007hyperbolicity} for  linearized Grad's moment systems.
While the results in \cite{ellis1975first,dudynski2013} are of perturbative nature (Taylor expansion in $k$), complete and explicit spectral analyses have been carried out by the present authors for the %three-component Grad equation \cite{kogelbauer2020slow}, 
the Bhatnagar--Gross--Krook (BGK) \cite{kogelbauerBGKspectral1} and the Shakhov \cite{kogelbauer2024spectral} kinetic models. In particular, critical wave numbers were explicitly evaluated for the BGK model in \cite{kogelbauerBGKspectral1}, the data referenced in Appendix \ref{app:BGK} to be used in the examples below.\\

%, see also \cite{cercignani1988boltzmann} for the calculation of spectral properties of quasi-equilibrium models using the Laplace transform.

%\begin{figure}
%    \centering
%    \includegraphics[width=0.8\linewidth]{diffcomp.png}
%    \caption{\footnotesize{Comparison of the exact diffusion mode obtained in \cite{kogelbauer2021} (solid line) to the numerical approximations obtained in \cite{karlin2014non} for the BGK equation. While local approximations, i.e., truncation's of the Chapman--Enskog series,  such as the Navier--Stokes extend unboundedly to all wave numbers, the hydrodynamic modes are only defined for a certain range of wave numbers.}}
%    \label{diffcomp}
%\end{figure}

\subsection{Spectral matrix and symmetries}
\label{sec:spec_matrix}

The hydrodynamic eigenvalues (and their associated eigenfunctions) are described through a $5\times 5$ \emph{spectral matrix},
\begin{equation}\label{GLP}
{{G}(\lambda,\bm{k})} = \int_{\mathbb{R}^3}d\mu(\bm{v}) {e}\otimes\Big((\mathcal{L}_{\bm{k}}+\mathbb{P}-\lambda)^{-1}{e}\Big).
\end{equation}
Indeed, since $\mathcal{Q}+\mathbb{P}$ is invertible, the eigenvector equation {$\mathcal{L}_{\bm{k}}f_{\bm{k}}=\lambda f_{\bm{k}}$} can be rewritten as a solution to the implicit equation, 
%\begin{equation}\label{flambdaimplicit}
{$f_{\bm{k}} = (\mathcal{L}_{\bm{k}}+\mathbb{P}-\lambda)^{-1}\mathbb{P}f_{\bm{k}}$}.
%\end{equation}
%for small wave number.
%by analytical spectral perturbation theory \cite{hislop2012introduction}. 
Applying $\mathbb{P}$,
%to \eqref{flambdaimplicit}, 
we see that the \textit{hydrodynamic moments},
\begin{equation}
    {{\eta}_{\bm{k}}(\lambda)=\mathbb{P} f_{\bm{k}}},
\end{equation}
of an eigenfunction associated with the eigenvalue $\lambda$ solve the equation,
\begin{equation}
 ({{G}(\lambda,\bm{k})}-{I}){{\eta_{\bm{k}}}(\lambda)}=0,
\end{equation}
where ${I}$ is unit matrix. Calculation of the spectral matrix \eqref{GLP} for quasi-equilibrium kinetic models such as BGK or Shakhov (finite-rank perturbation of the transport operator) is related to Weinstein--Aronszajn determinant \cite{kato1995perturbation,aronszajn1948rayleigh,WEINSTEIN1974604}, as detailed in \cite{kogelbauerBGKspectral1,kogelbauer2024spectral}.\\
 
%let us introduce a convenient coordinate system by splitting $\bm{v}=\bm{v}_{\parallel}+\bm{v}_{\perp}$ such that $\bm{v}_{\parallel}=\frac{1}{k^2}(\bm{v}\cdot\bm{k})\bm{k}$ and $ \bm{v}_{\perp}=-\frac{1}{k^2}\bm{k}\times(\bm{k}\times\bm{v})$, which satisfies $\bm{v}^{\perp}\cdot\bm{k} =0$. Let $\bm{Q}_{\bm{k}}$ denote the three-dimensional rotation matrix that realizes $\bm{v}=\bm{Q}_{\bm{k}}(v_{\parallel},v_{\perp 1},v_{\perp 2})$ together with $\bm{k}=\bm{Q}_{\bm{k}}(k,0,0)^T$ and define the $5\times 5$ block-diagonal matrix   $\tilde{Q}_{\bm{k}} = \diag(1,\bm{Q}_{\bm{k}},1)$.\\
The commutation property \eqref{commutation} allows us to infer a general structure of the spectral matrix. To see this, let us introduce a convenient coordinate system. For any $\bm{k}$, let $\bm{Q}_{\bm{k}}$ denote the unique rotation such that $\bm{k}=\bm{Q}_{\bm{k}}(k,0,0)$ and define the block-diagonal $5\times 5$ matrix,
\begin{equation}\label{eq:Qtilde}
    \tilde{{Q}}_{\bm{k}} = \diag(1,\bm{Q}_{\bm{k}},1).
\end{equation}  
Changing coordinates as  $\bm{v} = Q_{\bm{k}}\bm{w}$, $d\bm{v} = d\bm{w}$, in \eqref{GLP} and using the commutation property \eqref{commutation} we find that
%can write
%\begin{small}
%
%\begin{equation}\label{GL}
%{G}(\lambda) = \tilde{{Q}}_{\bm{k}}\int_{\bm{R}^3}d\mu(\bm{w}) {e}\otimes\Big((\mathcal{L}_{(k,0,0)}+\mathbb{P}-\lambda)^{-1}{e}\Big)\tilde{{Q}}_{\bm{k}}^T, 
%\end{equation}
%
%\end{small}
%
%where we have used $[\mathbb{P},U_{\bm{O}}]=0$. 
%Since $(\mathcal{L}_{(k,0,0)}+\mathbb{P}-\lambda)^{-1}$ commutes with reflections in $w_1$ and $w_2$, we conclude that 
%any odd entries in $w_2$ or $w_3$ appearing in \eqref{GL} will vanish as they are integrated against the even Gaussian measure. 
%Therefore, 
the spectral matrix $G$ \eqref{GLP} has the structure,
\begin{equation}\label{GLPast}
    {G}= \tilde{{Q}}_{\bm{k}}\begin{pmatrix}
        \ast & \ast & 0 & 0 & \ast\\
        \ast & \ast & 0 & 0 & \ast\\
        0 & 0 & \ast & 0 & 0\\
        0 & 0 & 0 & \ast  & 0\\
        \ast & \ast & 0 & 0 & \ast
    \end{pmatrix}\tilde{{Q}}_{\bm{k}}^T. 
\end{equation}
Proof of the structure of the spectral matrix \eqref{GLPast} is given in Appendix \ref{app:Gmatrix}.
We note in passing that a direct calculation of the spectral matrix for the BGK and Shakhov kinetic models \cite{kogelbauerBGKspectral1,kogelbauer2024spectral} confirm to the structure \eqref{GLPast}. Here, solely relying on the symmetry \eqref{commutation}, we prove the generic structure \eqref{GLPast} for any kinetic equation of the Boltzmann type.

\subsection{From spectral data to hydrodynamic equations}
\label{sec:spectraltohydro}

The final step of the general analysis is the transform from spectral data to the conventional hydrodynamic fields. For fixed $k$ and a given set of eigenvalues $\Lambda(k)$, the change from spectral coordinates to hydrodynamic variables is given by a finite-dimensional modification of the classical Riesz projection \cite{hislop2012introduction},
\begin{equation}\label{eq:Riesz}
	{H}(\Lambda) = -\frac{1}{2\pi\ri} \oint_{\Gamma(\Lambda)} \mathbb{P}(\mathcal{L}_{\bm{k}}-z)^{-1}\mathbb{P} \, dz,
\end{equation}
where $\Gamma(\Lambda)$ is any simply closed contour encircling the eigenvalues at wave number k once in positive direction, see Fig.\ \ref{spec1}.  

Evaluating the full Riesz projection in \eqref{eq:Riesz} would be equivalent to projecting onto the eigenvectors associated to $\Lambda(k)$.  We emphasize, however, that the $H$-matrix being defined as the Riesz-projection sandwiched between two finite-range $\mathbb{P}$-projections can be evaluated entirely from the (finite-dimensional) spectral matrix \eqref{GLPast} since the corresponding columns of $H$ are elements of the kernel of ${(G(\lambda,\bm{k})-I)}$. Since the diffusion mode and the pair of acoustic modes are simple, the corresponding kernel is one-dimensional. We also remark that, for general, linear collision kernels, formula \eqref{eq:Riesz} can be used as the basis for a complex-valued quadrature scheme for the computation of $H$.

Indeed, the structure of the spectral matrix \eqref{GLPast} immediately implies that 
the hydrodynamic moments associated to the shear mode (cf.\ third and fourth columns in \eqref{GLPast}) take the form,
\begin{equation}\label{basisshear}
\begin{split}
{{\eta}_{\bm{k}}}^{(1)}(\lambda_{\rm s})=\tilde{{Q}}_{\bm{k}}(0,0,1,0,0),\\
{{\eta}_{\bm{k}}}^{(2)}(\lambda_{\rm s})=\tilde{{Q}}_{\bm{k}}(0,0,0,1,0).
\end{split}
\end{equation}
%\begin{equation}\label{basisshear}\bm{\eta}_1(\lambda_{shear})=\tilde{Q}_{\bm{k}}\left(\begin{array}{c}0\\0\\1\\0\\0\end{array}\right),\qquad \bm{\eta}_2(\lambda_{shear})=\tilde{Q}_{\bm{k}}\left(\begin{array}{c}0\\0\\0\\1\\0\end{array}\right),\end{equation}
Furthermore,  the hydrodynamic moments associated to the diffusion and the acoustic modes are written as,
\begin{equation}\label{basissimple}
    {\eta_{\bm{k}}}(\lambda) = \tilde{{Q}}_{\bm{k}}\left(1 ,\frac{\ri\lambda}{k}, 0 , 0,{\theta(\lambda,\bm{k})}\right),\ \lambda\in\{\lambda_{\rm d},\lambda_{\rm a},\lambda_{\rm a}^{*}\},
\end{equation}
%\begin{equation}\label{basissimple}    \boldsymbol{\eta}(\lambda) = \tilde{Q}_{\bm{k}}\left(\begin{array}{c} 1 \\ \frac{\ri\lambda}{k}\\ 0 \\ 0\\ \theta(\lambda)\end{array}\right),\end{equation}
for a single function $\theta$, which we call \textit{spectral temperature}, evaluated on the eigenvalues and given by the quotient of Riesz projections, 
%which may be evaluated explicitly to 
\begin{equation}\label{eq:spectemp}
   %\theta(\lambda) = \frac{\oint_{\Gamma(\lambda)}[(\bm{G}(z)-\bm{I})^{-1}]_{5,5} dz}{\oint_{\Gamma(\lambda)}[(\bm{G}(z)-\bm{I})^{-1}]_{1,1} dz},
   {\theta(\lambda,\bm{k})} = \frac{\text{adj}[{G(\lambda,\bm{k})}-I]_{1,5}}{\text{adj}[{G(\lambda,\bm{k})}-I]_{1,1}}.
\end{equation}
Here $\text{adj}$ denotes adjugate matrix. The spectral temperature is an analytic function of the eigenvalue branch, depends only on wave number, $\theta(\lambda,\bm{k}) = \theta(\lambda,k)$, and  satisfies the commutation relation,
\begin{equation}\label{conjugatetheta}
    \theta(\lambda,k)^* = \theta(\lambda^*,k).
\end{equation}
Derivation of the formulas \eqref{basissimple}, \eqref{eq:spectemp} {and \eqref{conjugatetheta}} are given in Appendix \ref{app:Hmatrix}. {To ease notation, we suppress the $k$-dependence of $\theta$ below.}

We summarize at this point that, up to the rotation $\bm{Q}_{\bm{k}}$, the columns of Riesz projection matrix ${H}$ \eqref{eq:Riesz} are given by \eqref{basisshear} and $\eqref{basissimple}$, which takes the explicit form,
 \begin{equation}\label{defHfinal}
     H = \tilde{Q}_{\bm{k}} \begin{pmatrix}
         1 & 1 & 1 & 0 & 0\\
         \frac{\ri }{k}\lambda_{\rm d} &  \frac{\ri}{k}\lambda_{\rm a} &  \frac{\ri}{k}\lambda_{\rm a}^* & 0 & 0\\
         0 & 0 & 0 & 1 & 0\\
         0 & 0 & 0 & 0 & 1\\
         \theta(\lambda_{\rm d}) &  \theta(\lambda_{\rm a}) &  \theta(\lambda_{\rm a}^*) & 0 & 0
     \end{pmatrix}.
 \end{equation}
%, and evaluated at \eqref{LambdaHydro}. 
This general result for the transformation matrix $H$ is consistent with a direct computation for the BGK model by evaluating the contour integral \eqref{eq:Riesz} \cite{kogelbauerBGKspectral2}.

Let us denote the Fourier-images of the hydrodynamic variables in $\bm{Q}_{\bm{k}}$-coordinates as 
\begin{equation}\label{eq:primitive}
\hat{{h}} = (\hat{\rho},\hat{u}_{\parallel},\hat{u}_{\perp 1},\hat{u}_{\perp 2},\hat{T}).\end{equation} 
With the matrix of eigenvectors associated to \eqref{LambdaHydro},  
\begin{equation}
	F_{{\Lambda}} = \diag(\hat{f}_{\lambda_{\rm s},1},\hat{f}_{\lambda_{\rm s},2}, \hat{f}_{\lambda_{\rm d}},\hat{f}_{\lambda_{\rm a}},\hat{f}_{\lambda_{\rm a}^*}),
\end{equation}
the spectral closure operator, $\mathcal{C}\hat{{h}}  = \mathbb{P}^{\perp}F_{{\Lambda}}{H}^{-1}\hat{{h}}$, verifies the aforementioned invariance equation \eqref{eq:invariance} while the hydrodynamic equations are written as,
\begin{equation}\label{dynh}
    \partial_t \hat{{h}}= \mathcal{T} \hat{{h}},\ \mathcal{T} = {H}\tilde{\Lambda} {H}^{-1}, 
\end{equation}
where $\tilde{\Lambda} = \diag(\Lambda)$ and $\det{H}\neq 0$ is guaranteed for small wave numbers by continuity. 
Thanks to the stability of the spectrum, the linear system \eqref{dynh} is automatically hyperbolic. It provides a non-perturbative and unique closure compared to the regularization of higher-order hydrodynamics \cite{bobylev2020kinetic}.
With the structural properties \eqref{basisshear}, \eqref{basissimple} and \eqref{defHfinal}, the transport matrix $\mathcal{T}$ becomes,
\begin{equation}\label{defT}
    \mathcal{T} = \begin{pmatrix}
0 & -\ri k & 0 & 0 & 0\\
\ri \tau_1 &  \tau_2 & 0 & 0 & \ri \tau_3\\
0 & 0 & \lambda_{\rm s} & 0 & 0\\
0 & 0 & 0 & \lambda_{\rm s} & 0\\
\tau_4 & \ri \tau_5 & 0 & 0 & \tau_6
    \end{pmatrix},
\end{equation}
where the generalized transport coefficients are given explicitly in terms of eigenvalues and spectral temperature,
{on the account of the symmetry \eqref{conjugatetheta},}
%\begin{widetext}
%\begin{small}
\begin{equation}\label{transportcoef}
    \begin{split}
 %    \tau_1 & =\frac{2}{k^2\det\bm{H}_{\bm{k}}}\Big[ \lambda_{\rm diff}\Im[\lambda_{\rm ac}^*(\lambda_{\rm diff}-\lambda_{\rm ac})\theta(\lambda_{\rm ac})]-|\lambda_{\rm ac}|^2(\Im\lambda_{\rm ac})\theta(\lambda_{\rm diff})  \Big],\\
     %
&     \tau_1  =\frac{2}{k^2\det{H}}\left[ \lambda_{\rm d}\Im[\lambda_{\rm a}^*(\lambda_{\rm d}-\lambda_{\rm a})\theta(\lambda_{\rm a})]-|\lambda_{\rm a}|^2(\Im\lambda_{\rm a})\theta(\lambda_{\rm d})  \right],\\
     %
 %     \tau_2 & =  -\frac{2}{k\det\bm{H}_{\bm{k}}}\Big[\Im[(\lambda_{\rm diff}^2-(\lambda_{\rm ac}^*)^2)\theta(\lambda_{\rm ac})]-2(\Re\lambda_{\rm ac})(\Im\lambda_{\rm ac})\theta(\lambda_{\rm diff})\Big],\\
      %
&      \tau_2  =  \frac{2}{k\det{H}}\left[2(\Re\lambda_{\rm a})(\Im\lambda_{\rm a})\theta(\lambda_{\rm d})-\Im[(\lambda_{\rm d}^2-(\lambda_{\rm a}^*)^2)\theta(\lambda_{\rm a})]\right],\\
      %
 %     \tau_3 & = \sqrt{\frac{3}{2}}\frac{2}{k^2\det\bm{H}_{\bm{k}}}|\lambda_{\rm diff}-\lambda_{\rm ac}|^2(\Im\lambda_{\rm ac}),\\
      %
&      \tau_3  = \frac{\sqrt{6}}{k^2\det{H}}|\lambda_{\rm d}-\lambda_{\rm a}|^2(\Im\lambda_{\rm a}),\\
  %
%    \tau_4 & = \sqrt{\frac{3}{2}}\frac{2}{k\det\bm{H}_{\bm{k}}} \Big[\Im[\lambda_{\rm ac}(\lambda_{\rm diff}-\lambda_{\rm ac})\theta(\lambda_{\rm ac})\theta(\lambda_{\rm diff})]+\lambda_{\rm diff}(\Im\lambda_{\rm ac})|\theta(\lambda_{\rm ac})|^2  \Big],\\
%
& \tau_4  = \frac{\sqrt{6}}{k\det{H}} \left[\Im[\lambda_{\rm a}(\lambda_{\rm d}-\lambda_{\rm a})\theta(\lambda_{\rm a})\theta(\lambda_{\rm d})]+\lambda_{\rm d}(\Im\lambda_{\rm a})|\theta(\lambda_{\rm a})|^2  \right],\\
%
%     \tau_5 & = \sqrt{\frac{3}{2}}\frac{2}{\det\bm{H}_{\bm{k}}} \Big[\theta(\lambda_{\rm diff})\Im[\theta(\lambda_{ac})(\lambda_{\rm diff}-\lambda_{\rm ac})]+(\Im\lambda_{\rm ac})|\theta(\lambda_{ac})|^2 \Big],\\
%
& \tau_5  = \frac{\sqrt{6}}{\det{H}} \left[\theta(\lambda_{\rm d})\Im[\theta(\lambda_{a})(\lambda_{\rm d}-\lambda_{\rm a})]+(\Im\lambda_{\rm a})|\theta(\lambda_{a})|^2 \right],\\
%
%     \tau_6  & = \frac{2}{k\det\bm{H}_{\bm{k}}} \Big[\lambda_{\rm diff}\theta(\lambda_{\rm diff})(\Im\lambda_{\rm ac})+\Im[\lambda_{\rm ac}\theta(\lambda_{\rm ac})(\lambda_{\rm ac}^*-\lambda_{\rm diff})] \Big].
%
& \tau_6   = \frac{2}{k\det{H}} \left[\lambda_{\rm d}\theta(\lambda_{\rm d})(\Im\lambda_{\rm a})+\Im[\lambda_{\rm a}\theta(\lambda_{\rm a})(\lambda_{\rm a}^*-\lambda_{\rm d})] \right],  \\
&\det{H} = \frac{2}{k}\left((\Im\lambda_{\rm a})\theta(\lambda_{\rm d})-\Im[(\lambda_{\rm a} - \lambda_{\rm d})\theta(\lambda_{\rm a}^*)]\right).    
    \end{split}
    \end{equation}
%   \end{small}
%\end{widetext}
%and $$. 
Since the generalized transport coefficients \eqref{transportcoef} are algebraic expressions of the eigenvalue branches \eqref{LambdaHydro}, they only exist up to the first critical wave number, which, in turn, depends on the collision operator.
The rigorous map of spectral variables onto the hydrodynamic equations in conventional primitive variables is our first main result. This spectral calculation rigorously derives the checker-board structure of the transport matrix that was observed for special cases by different techniques, e.\ g.\ in \cite{karlin2002grad,PhysRevLett.100.214503,colangeli2009boltzmann}.
We shall now proceed with select implications of these new hydrodynamics.

\section{Slemrod's conjecture and viscosity-capillarity balance}
\label{sec:korteweg}

%\onecolumngrid

\begin{figure}[h]
\includegraphics[width=0.9\linewidth]{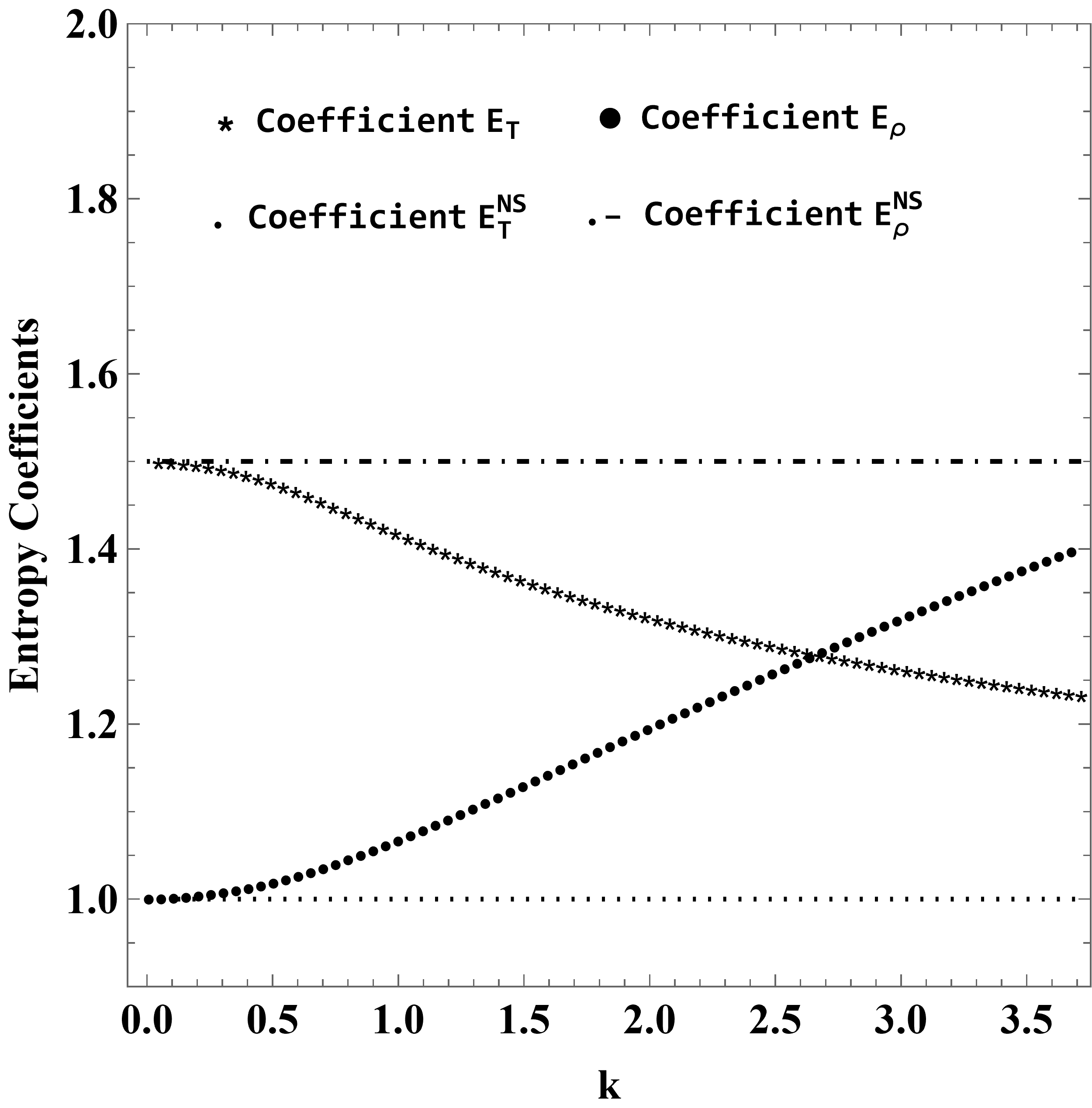}
 \caption{\footnotesize{Comparison of the dissipation coefficient of the Navier--Stokes equation (dashed) to the dissipation coefficient obtained from the BGK hydrodynamics ($*$-symbol).} }
 \label{compE}
\end{figure}

\begin{figure}
    \centering
    \includegraphics[width=0.9\linewidth]{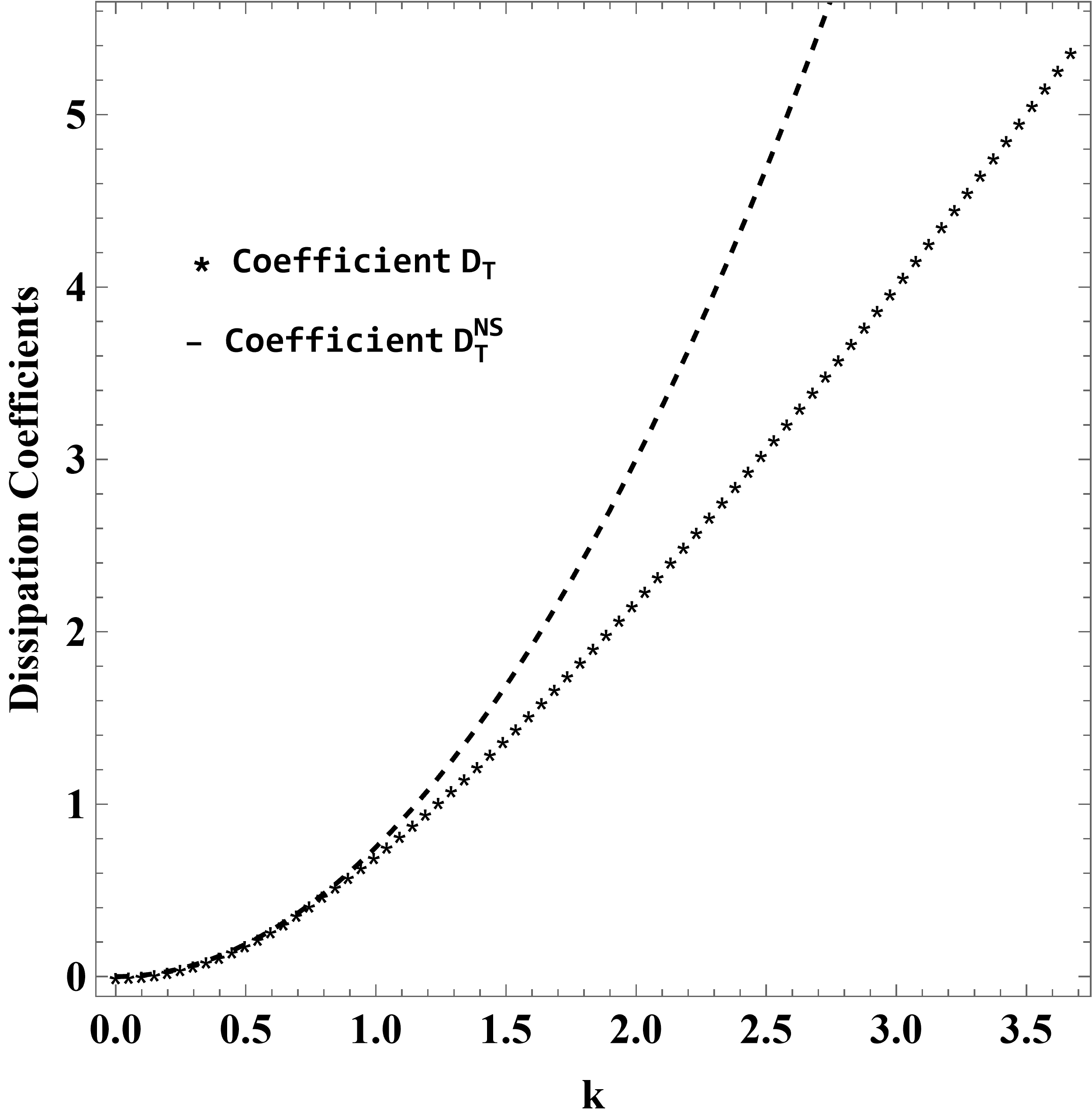}
    \caption{\footnotesize{Comparison of the entropy coefficients of the Navier--Stokes equation (dashed-dotted and dotted) to the entropy coefficient obtained from the BGK hydrodynamics ($*$-symbol and big dots).} }
    \label{compD}
\end{figure}
%\twocolumngrid

%\textcolor{blue}{From the construction of the spectrally-closed hydrodynamics in the previous section \eqref{dynh},  }

As exemplified on the exact summation of the Chapman--Enskog series for Grad's moment system \cite{gorban1996short}, the full hydrodynamic closure induces a modification of entropy that ensures pure dissipation in the hydrodynamic variables \cite{slemrod2012chapman}. This can be regarded as a non-local extension of Korteweg's theory of capillarity \cite{1573105975105495040}. It was conjectured by Slemrod \cite{SLEMROD20131497} that this feature persists for exact hydrodynamics derived from general kinetic models.
Hence, guided by Onsager's principle \cite{onsager1931reciprocal1,onsager1931reciprocal2}, we seek a physical coordinate system for which a hydrodynamic entropy matrix ${E}_{\rm hydro}$ and a modified dissipation matrix ${D}_{\rm hydro}$ are both diagonal and positive semi-definite such that the entropy-dissipation relation,

\begin{equation}\label{entropydiss}
    {\partial_t} {E}_{\rm hydro}\tilde{{h}}\cdot\tilde{{h}} ^* = - {D}_{\rm hydro}\tilde{{h}}\cdot\tilde{{h}}^*,
\end{equation}

holds for some linear combination of the basic hydrodynamic variables,
\begin{equation}
    \tilde{h} = V\hat{h},
\end{equation}
for some invertible matrix $V$ to be determined. Defining $\tilde{\mathcal{T}}= V\mathcal{T}V^{-1} $, the combined moments satisfy $\partial_t\tilde{h} =\tilde{\mathcal{T}}\tilde{h}$.\\
Any two such matrices are necessarily related through the continuous Lyapunov equation, 
\begin{equation}\label{eq:lyapunov}
{\tilde{\mathcal{T}}}^*{E}_{\rm hydro}+{E}_{\rm hydro}{\tilde{\mathcal{T}}}=-{D}_{\rm hydro}.
\end{equation} 
Indeed, writing out the time derivative in \eqref{entropydiss} explicitly and using the dynamics \eqref{dynh} together with the symmetry of $E_{\rm hydro}$ gives
\begin{equation}
    \begin{split}
        {\partial_t} {E}_{\rm hydro}\tilde{{h}}\cdot\tilde{{h}}^* & = {E}_{\rm hydro}{\partial_t\tilde{{h}}}\cdot\tilde{{h}}^* + {E}_{\rm hydro}\tilde{{h}}\cdot \partial_t\tilde{{h}}^*\\
         & = \tilde{{h}}\cdot (E_{\rm hydro}{\tilde{\mathcal{T}}} + {\tilde{\mathcal{T}}}^* E_{\rm hydro})\tilde{{h}}^*,
    \end{split}
\end{equation}
which, in order to hold for all $\tilde{h}$, is equivalent to \eqref{eq:lyapunov}. \\
Note that the algebraic structure of the transport matrix \eqref{defT} prevents an entropy-dissipation balance of the form \eqref{entropydiss} for the original hydrodynamic variables $\hat{{h}}$ \eqref{eq:primitive}. Indeed, assuming a diagonal dissipation matrix $D_{\rm hydro}$ for $\tilde{h}=\hat{h}$, i.e., $V=I$, the only solution to the Lyapunov equation \eqref{eq:lyapunov} implies $E_{\rm hydro}=D_{\rm hydro}=0$ (for $k\neq 0$ for which $\tilde{\mathcal{T}}\neq 0$).\\
However, one off-diagonal linear combination of basic hydrodynamic variables suffices to guarantee the desired balance law. 
Up to this re-definition of hydrodynamic variables, the balance law \eqref{entropydiss} is then unique. Indeed, for the change of coordinates,
\begin{equation}\label{V}
    V = \begin{pmatrix}
        1 & 0 & 0 \\
        0 & \bm{I} & 0\\
        \frac{\tau_4}{\tau_6} & 0 & 1
    \end{pmatrix},
\end{equation}
which defines a (non-locally) modified temperature,
\begin{equation}
    \tilde{T}=\hat{T}+\frac{\tau_4}{\tau_6}\hat{\rho},
\end{equation}
we obtain,
\begin{equation}\label{entdiss}
\begin{split}
{\partial_t}\left(E_{\rho}|\hat{\rho}|^2+|\hat{\bm{u}}|^2+{E_{T}}|{\tilde{T}}|^2\right)&\\
 = -2\tau_2 |\hat{u}_{\parallel}|^2-\lambda_{\rm s}|\hat{\mathbf{u}}_{\perp}|^2&-2{D_{T}} |{\tilde{T}}|^2,
\end{split}
\end{equation}
where the entropy coefficients and the dissipation coefficient are given by,
\begin{equation}\label{entropydissipation}
    E_{\rho} = \frac{\tau_3\tau_4-\tau_1\tau_6}{k \tau_6},\ {E_T =\frac{\tau_3\tau_6}{\tau_5\tau_6-k\tau_4}},\ {D_{T}} = \tau_6 E_p.
\end{equation}
The coordinate system \eqref{V} is minimal in the sense that it only contains one off-diagonal entry to ensure the existence of diagonal solutions $(E_{\rm hydro},D_{\rm hydro})$ of the Lyapunov equation \eqref{eq:lyapunov}. Clearly, the entropy dissipation relation is only defined up to rescaling of the variables and there might exists other coordinate systems for which $(E_{\rm hydro},D_{\rm hydro})$ are still positive semi-definite but not diagonal.\\
We emphasize that the entropy-dissipation coefficients \eqref{entropydissipation} depend on the eigenvalue branches \eqref{LambdaHydro} through the generalized transport coefficients \eqref{transportcoef} and thus only exist up to the first critical wave number.
The hydrodynamic variables in \eqref{entropydiss} become,
\begin{equation}
    \tilde{h} = (\hat{\rho},\hat{\bm{u}},\tilde{T}),\quad \hat{\bm{u}} = (\hat{u}_{\parallel},\hat{\bm{u}}_{\perp}),
\end{equation}
while the hydrodynamic entropy and dissipation matrix take the explicit form,
\begin{equation}
    E_{\rm hydro} = \begin{pmatrix}
        E_{\rho} & 0 & 0 \\
        0 & \bm{I} & 0\\
        0 & 0 & E_{T}
    \end{pmatrix},\  D_{\rm hydro} = \begin{pmatrix}
        0 & 0 & 0\\
        0 & \bm{D}_u & 0 \\
        0 & 0 & 2D_{T}
    \end{pmatrix},
\end{equation}
with the diagonal matrix $\bm{D}_u = \diag(2\tau_2,\lambda_s,\lambda_s)$. \\
Thus, the entropy-dissipation relation \eqref{entdiss} proves Slemrod's conjecture for a general kinetic equation.
We compare the $k$-dependent coefficients \eqref{entropydissipation} to those of the Navier--Stokes equation, 
\begin{equation}
    E_{\rho}^{\rm NS} = 1,\ {E_{T}^{\rm NS}} = {3}/{2},\ {D_T^{\rm NS}} = ({3}/{2})\mu k^2,
\end{equation}
where $\mu$ is the thermal conductivity \cite{chapman1990mathematical}. Figure \ref{compD} and Figure \ref{compE} show a comparison of the entropy and dissipation coefficients of the Navier--Stokes equation and the BGK hydrodynamics.
Interestingly, a controversial issue about thermodynamics of hydrodynamic systems with higher derivatives, see e.\ g.\ \cite{PhysRevA.31.2502}, is resolved as an immediate consequence of the present approach through the balance law \eqref{entdiss}. Constraining the original system \eqref{maineq} by passing to the slow manifold associated to \eqref{LambdaHydro} thus induces a non-local, wave-number-dependent modification of the entropy.\\
We note that the change of coordinates \eqref{V} that puts the non-local hydrodynamic equations into entropy-dissipation form is reminiscent of the procedures carried out in \cite{slemrod2012chapman,colangeli2007hyperbolicity} to obtain an $H$-theorem for certain model problems. Indeed, the local truncations of of the Chapman--Enskog expansion can be made hyperbolic by a change of coordinates, called Bobylev regularization \cite{bobylev2006instabilities,bobylev2020kinetic}. \\
Finally, we remark that the Lyapunov equation \eqref{eq:lyapunov} also appears in the context of fluctuation-dissipation theory \cite{zwanzig2001nonequilibrium}, where it links the diffusion matrix, the transport matrix and the stationary covariance matrix. It might be interesting to exploit this connection with non-local hydrodynamics further.

\begin{comment}
    \begin{figure}   
\centering 
\includegraphics[width=0.8\linewidth]{figErhocomp.png}   
\caption{The entropy-dissipation coefficients \eqref{entropydissipation} calculated for the linear BGK equation with relaxation time $\tau=0.3$ compared to the asymptotic approximation \eqref{EDapprox} (dashed line) and first critical wave number $k_{\rm crit, min}= 3.72$ .}    
\label{figErho}
\end{figure}
\end{comment}

\section{Knudsen paradox}
\label{sec:knudsen}

To that end, results presented above are valid for any linear kinetic equation of Boltzmann type. Further analysis of the new hydrodynamic equations only depends on the explicit information about the spectrum. In the remainder of this paper, we use closed-form results for the spectrum of the BGK model \cite{kogelbauerBGKspectral1}. For convenience, a summary of spectral properties of the linearized BGK model are provided in Appendix \ref{app:BGK}.

%First, we compare the full non-local entropy-dissipation coefficients \eqref{entropydissipation} to the Navier--Stokes approximation\begin{equation}\label{EDapprox}E_{\rho}^{\rm NS} = 1,\quad E_{p}^{\rm NS} = \frac{3}{2},\quad D_p^{\rm NS} = \frac{5}{2}\tau k^2,\end{equation}\textcolor{blue}{where $\tau$ is the conventional BGK relaxation parameter.} As an example, the coefficients \eqref{entropydissipation} and \eqref{EDapprox} for the BGK operator are shown in Fig.\ \ref{figErho}. The dissipation coefficient on the Navier--Stokes level implies stronger decay properties as compared to the hydrodynamics on the BGK level. 

\begin{figure}    \centering    \includegraphics[width=1\linewidth]{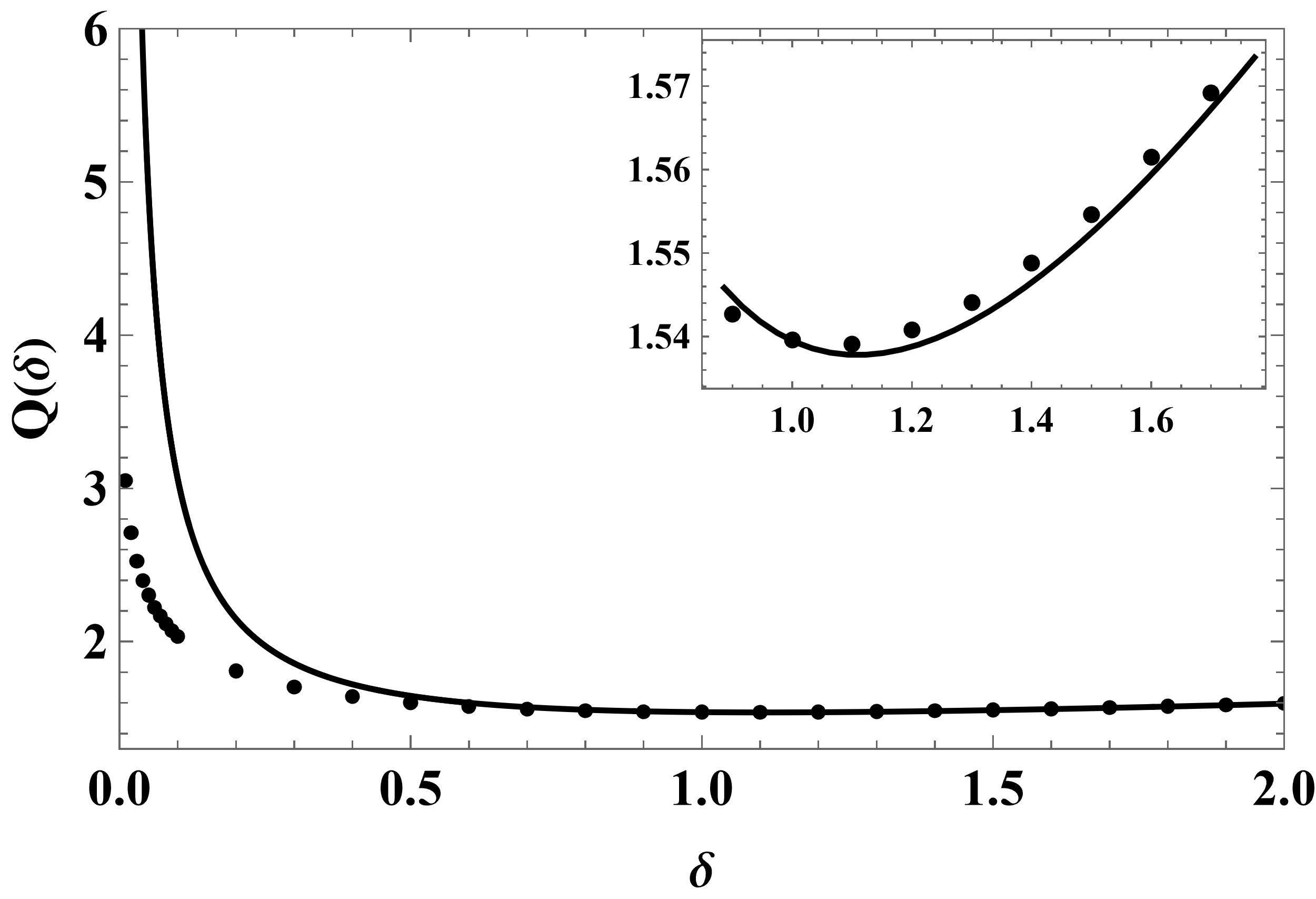}
\caption{\footnotesize{Normalized flow rate in a channel flow $Q$ \eqref{defQ} as a function of inverse Knudsen number $\delta$. Symbol: Linearized BGK model, Cercignani \& Daneri \cite{cercignani1963flow};
Solid line: Present hydrodynamic spectral closure \eqref{Qshear} and \eqref{shear}. Inset: comparison close to the Knudsen minimum.}}   
%\caption{Comparison of the normalized flow rate obtained in \cite{cercignani1963flow} from the BGK equation (symbol) compared to the normalized flow rate obtained from the shear mode in the spectral closure \eqref{Qshear} (solid line). }    
\label{figKnudesenMinimum}\end{figure}

We test the hydrodynamic equations \eqref{dynh} and \eqref{defT} on a classical benchmark problem - the existence of Knudsen minimum in a channel flows \cite{akhlaghi2023comprehensive,porodnov1974experimental}. Consider a long channel with rectangular cross-section and width $d$ ($x$-direction). Following the set-up of Cercignani \& Daneri \cite{cercignani1963flow}, we assume a constant density, isothermal stationary flow of a gas induced by a uniform pressure drop $\Delta p$ along the channel with no-slip boundary conditions for the linearized BGK model. The normalized flow rate is,
\begin{equation}\label{defQ}
    Q(\delta) = -\frac{\rho}{d^2\Delta p} \int_{-{d}/{2}}^{{d}/{2}} dx\,  u_z(x),
\end{equation}
where $u_z$ is the velocity of the gas in the $z$-direction and $\delta=d/\tau $ is the inverse Knudsen number for the BGK relaxation time $\tau$ \cite{cercignani1963flow}. 
%Derived from the BGK equation, 
%with the relaxation time $\tau$, 
With a simple Fourier analysis, the transport matrix \eqref{defT} allows us to evaluate \eqref{defQ} to
\begin{equation}\label{Qshear}
    Q^{\rm hydro}(\delta) =  u_{z,0} - \frac{4}{\pi^3d}\sum_{j=0}^\infty \frac{1}{(2j+1)^2 \lambda_{\rm s}(2j+1)}, 
\end{equation}
 where the constant base flow is given explicitly by
 \begin{equation}
     u_{z,0}=\sum_{j=0}^{\infty}\frac{1}{(2j+1)^{4}}\approx 1.0146. 
 \end{equation}
The result \eqref{Qshear} is valid for any collision model, details of the derivation are presented in Appendix \ref{app:Kn}. Further evaluation concerns the BGK model, for which 
 the shear mode takes explicit form \cite{kogelbauerBGKspectral1}:
\begin{equation}\label{shear}
    \lambda_s(k)  = -k\frac{\omega_x}{2}\ri  Z^{-1}\left(\ri k \frac{\pi}{\delta}\right)-\frac{1}{2\pi} \omega_x\delta,
\end{equation}
where $Z$ is the plasma dispersion function (\cite{fried2015plasma}, see Eq.\ \eqref{defZ} in Appendix \ref{app:BGK}), $k$ is the wave number and 
\begin{equation}
    \omega_x=\frac{2\pi}{d}
\end{equation}
is the basic frequency of the channel. The mode \eqref{shear} is extended as $\lambda_s=-{1}/{\tau}$ beyond its critical wave number,
\begin{equation}
    k_{\rm crit, s}=\sqrt{\frac{\pi}{2}}\frac{1}{\tau},
\end{equation}
see \cite{kogelbauerBGKspectral1} and Appendix \ref{app:BGK}.

Fig.\ \ref{figKnudesenMinimum} shows the dependence of the flow rate \eqref{defQ} on the inverse Knudsen number  \cite{cercignani1963flow}.
While \eqref{Qshear} is fully consistent with the small Knudsen number limit $\delta\to\infty$ \cite{cercignani1963flow},
%, $Q^{\rm BGK}\sim 1.0161+\delta/6$ for $\delta\to\infty$, 
the asymptotics at large Knudsen numbers,
\begin{equation}
    Q^{\rm hydro}\sim 1.0146 +4/\pi^3 \delta^{-1},\quad \delta\to 0,
\end{equation}
equation \eqref{Qshear} 
%$Q^{\rm hydro}\sim  \delta^{-1}$ \eqref{Qshear} 
%$Q^{\rm hydro}\sim 1.0146 +4/\pi^3 \delta^{-1}$ \eqref{Qshear} 
over-predicts relative to 
$Q^{\rm BGK}\sim -\pi^{-\frac{1}{2}}\log\delta$, see
%$Q^{\rm BGK}\sim -\log\delta$ 
\cite{cercignani1963flow}.
%$Q^{\rm BGK}\sim -\pi^{-\frac{1}{2}}\log\delta$ 
%for large Knudsen numbers $\delta\to 0$ \cite{cercignani1963flow}. 
Nevertheless, the hydrodynamic modes \eqref{shear} remain essential even for a substantial rarefaction, up to at least $\delta\sim 0.5$, while the Knudsen minimum is captured accurately:
For the full BGK model, the minimal flow rate is derived in \cite{cercignani1963flow},
\begin{equation}
   Q_{\rm min}^{\rm BGK}\approx 1.5391 \ \text{ at }  \ \delta_{\rm min}^{\rm BGK}\approx 1.1,
\end{equation}
while the minimal flow rate from \eqref{Qshear} yields,
\begin{equation}
    Q_{\rm min}^{\rm hydro}\approx 1.5378\ \text{ at }\ \delta_{\rm min}^{\rm hydro}\approx 1.1225,
\end{equation} 
thus showing excellent agreement.
We note that a similar asymptotics was obtained in \cite{struchtrup2007h} for the linear regularized thirteen moments Grad system \cite{karlin1998dynamic,struchtrup2003regularization}. 
Apart from this quantitative comparison, the present hydrodynamic theory also provides  a qualitative explanation of the Knudsen minimum:
%We emphasize that the existence of a critical wave number is crucial for the existence of the Knudsen minimum: 
While the shear mode of conventional local fluid models such as the Navier--Stokes or Burnett approximations decreases unboundedly with the increase of wave number, the exact shear mode of the spectral closure remains bounded due to the existence of the critical wave number. Hence, the series \eqref{Qshear} becomes divergent for $\delta\to 0$ thus implying a minimum of the flow rate at a finite Knudsen number.

\section{Conclusion}
\label{sec:conclusion}
We have presented a comprehensive solution to the problem of deriving near-equilibrium hydrodynamics from linear kinetic equations without any smallness assumptions on Knudsen number. The rigorous result of this article, the transport matrix in terms of spectral data, provides a universal starting point for non-local linear hydrodynamics. The derivation is solely based on the symmetry property of the underlying kinetic equation and the projection onto the slow manifold. In particular, it does not involve any ad-hoc assumptions on the dependence on higher-order moments thus rendering it the dynamically optimal solution: any other potential closure will deviate more from the actual kinetic solution. \\
The newly derived non-local hydrodynamics are capable of capturing rarefaction effects, of which the Knudsen minimum is one example elaborated in this paper. Since the hydrodynamic equations are entirely formulated in terms of dissipation relations, they are able to incorporate a variety of boundary conditions (as we shall demonstrate in forthcoming publications).

%Finally, the non-linear extension of the present theory shall be based on the deformation of the linear slow invariant manifold, which is the subject of our future work.

%Let us conclude with a summary of the results: We started out with a general linear kinetic equation including a transport and a collision term \eqref{maineq}, such as the Boltzmann equation linearized around a global Maxwellian. Using a commutation property in frequency space \eqref{commutation} in combination with the hydrodynamic spectrum \eqref{LambdaHydro}, we first derived the general structure of the spectral matrix \eqref{GLPast} and finally arrived at a closed-form expression of the transport coefficients in terms of spectral quantities \eqref{transportcoef} - the first main result of this paper. 
%Based on the properties of the transport matrix \eqref{defT}, we obtained a non-local entropy-dissipation balance \eqref{entdiss} and \eqref{entropydissipation}, generalizing Korteweg's theory of capillarity - the second main result of this paper.
%which was then compared to the Navier--Stokes equation \eqref{EDapprox}. 

\begin{acknowledgments}
This work was supported by the European Research Council (ERC) Advanced Grant  834763-PonD. 
Computational resources at the Swiss National Super  Computing  Center  CSCS  were  provided  under the grant s1286.
\end{acknowledgments}

\bibliographystyle{apsrev4-1}
\bibliography{DynamicalSystems}

\clearpage

\appendix

\section{Commutation Property of the Boltzmann Kinetic Operator}
\label{app:commprop}
The linearized Boltzmann collision integral takes the form
%\begin{small}
\begin{equation}\label{defLQ}
\begin{split} 
\mathcal{Q} = \int_{\mathbb{S}^2}\int_{\mathbb{R}^3}d\mu(\bm{v}_{*}) d\bm{\omega} B(\bm{v}-\bm{v}_{*},\boldsymbol{\omega})\\\times[f(\bm{v}_{*}')+f(\bm{v}')-f(\bm{v}_{*})-f(\bm{v})],
\end{split}
\end{equation}
%\end{small}
where the pre- and post-collisional velocities are given as $\bm{v}' =\bm{v}-(\bm{v}-\bm{v}_{*})\cdot\boldsymbol{\omega} \boldsymbol{\omega}$, $\bm{v}'_{*} =\bm{v}_{*}+(\bm{v}-\bm{v}_{*})\cdot\boldsymbol{\omega}  \boldsymbol{\omega}$, and the collision kernel $B$ depends on $\bm{v}-\bm{v}_*$ and $\omega$ only through $|\bm{v}-\bm{v}_*|$ and the deflection angle $\chi$ between $\bm{v}-\bm{v}_*$ and $\bm{v}'-\bm{v}_*'$, 
\begin{equation}
    B(\bm{v}-\bm{v}_{*},\boldsymbol{\omega}) = B(|\bm{v}-\bm{v}_{*}|,\chi(\bm{v}-\bm{v}_{*})).
\end{equation}
In order to show that $(-\ri\bm{k}\cdot\bm{v} -\mathcal{Q})$ satisfies the commutation property \eqref{commutation}, we first note that 
\begin{equation}\label{defw}   
\begin{split}       
-\ri\bm{k}\cdot \bm{v}f(O\bm{v}) & = -\ri O\bm{k}\cdot O\bm{v}f(O\bm{v}) \\ 
& = (U_O\mathcal{S}_{O\bm{k}}f)(\bm{v}),    
\end{split}   
\end{equation}
proving that the transport operator $(-\ri\bm{k}\cdot\bm{v})$ satisfies \eqref{commutation}. The transformation
\begin{equation} 
\bm{w} =  O\bm{v},\quad \bm{w}_* =  O\bm{v}_*,\quad \tilde{\boldsymbol{\omega}}=O\boldsymbol{\omega},    \end{equation}
leaves the pre- and post-collisional velocities invariant,
\begin{equation}  
\begin{split}      
\bm{w}' & =\bm{w}-(\bm{w}-\bm{w}_{*})\cdot\tilde{\boldsymbol{\omega}}  \tilde{\boldsymbol{\omega}},\\    \bm{w}'_{*} & =\bm{w}_{*}+ (\bm{w}-\bm{w}_{*})\cdot\tilde{\boldsymbol{\omega}}  \tilde{\boldsymbol{\omega}},        \end{split}   
\end{equation}  
and does not alter the volume elements in \eqref{defLQ}. Consequently, the deflection angle $\chi = \angle(\bm{v}-\bm{v}_*,\bm{v}'-\bm{v}_*') = \angle(\bm{w}-\bm{w}_*,\bm{w}'-\bm{w}_*')$ and, hence, the collision kernel is invariant under \eqref{defw} as well and we conclude that 
\begin{equation}       
[U_O,\mathcal{Q}] = 0.    
\end{equation}
Finally, let us note that
\begin{equation}  
[U_O,\mathbb{P}] = 0, 
\end{equation}
as well since 
\[U_O\mathbb{P}f=\langle f,e\rangle \cdot e(O\bm{v}) = \langle f(\bm{O}\bm{v}),e\rangle \cdot e(\bm{v}) = \mathbb{P}U_Of.\]
The commutation property \eqref{commutation} immediately carries over to the resolvent used in the definition \eqref{GLP}, thus allowing for the simplification carried out in Appendix B.   

\section{Structural properties of the eigenvectors and spectral matrix }
\label{app:Gmatrix}
We derive the general form of the hydrodynamic moments of an eigenvector based on the commutation property \eqref{commutation}. Rewriting the integral expression \eqref{GLP} in the $\bm{k}$-aligned coordinate system, $\bm{v} = \bm{Q}_{\bm{k}}\bm{w}$, $d\bm{v} = d\bm{w}$, and using \eqref{commutation} we find that,
%\begin{small}
\begin{equation}    
\begin{split}   
& \int_{\mathbb{R}^3}d\mu(\bm{v}) {e}(\bm{v})\otimes\Big((\mathcal{L}_{\bm{k}}+\mathbb{P}-\lambda)^{-1}{e}(\bm{v})\Big)\\      
&\quad = \int_{\mathbb{R}^3}d\mu(\bm{w}) U_{\bm{Q}_{\bm{k}}}{e}({\bm{w}})\otimes\Big(U_{\bm{Q}_{\bm{k}}}(\mathcal{L}_{\bm{k}}+\mathbb{P}-\lambda)^{-1}{e}(\bm{w})\Big)\\       
&\quad  =\int_{\mathbb{R}^3}d\mu(\bm{w})U_{\bm{Q}_{\bm{k}}} {e}({\bm{w}})\otimes\Big((\mathcal{L}_{\bm{Q}_{\bm{k}}^T\bm{k}}+\mathbb{P}-\lambda)^{-1}U_{\bm{Q}_{\bm{k}}}{e}(\bm{w})\Big).    
\end{split}
\end{equation}
%\end{small}
Since $\bm{Q}_{\bm{k}}^T\bm{k}=(k,0,0)$ and ${e}(\bm{Q}_{\bm{k}}\bm{w})= \tilde{Q}_{\bm{k}}{e}(\bm{w})$, it follows that   \begin{equation}
\begin{split}
(\mathcal{L}_{\bm{Q}_{\bm{k}}^T\bm{k}}+\mathbb{P}&-\lambda)^{-1}U_{\bm{Q}_{\bm{k}}}{e}(\bm{w})\\& = (\mathcal{L}_{(k,0,0)}+\mathbb{P}-\lambda)^{-1}\tilde{Q}_{\bm{k}}{e}(\bm{w})\\   
& = \tilde{Q}_{\bm{k}}(\mathcal{L}_{(k,0,0)}+\mathbb{P}-\lambda)^{-1}{e}(\bm{w}),     
\end{split}
\end{equation}
and we can write,
{
%\begin{small}
\begin{equation}
\label{GL}
{G(\lambda,\bm{k})} = \tilde{Q}_{\bm{k}} G_{\rm red}(\lambda,k)  \tilde{Q}_{\bm{k}}^T,
\end{equation}
for the reduced spectral matrix $G_{\rm red}$,
\begin{equation}
    G_{\rm red}(\lambda,k) = \int_{\mathbb{R}^3}d\mu(\bm{w}) e(\bm{w})\otimes\left((\mathcal{L}_{(k,0,0)}+\mathbb{P}-\lambda)^{-1}e(\bm{w})\right).
\end{equation}
}
%\end{small}
%\todo[inline]{notation needs to be consistent with main text}
%\todo[inline, color = green]{Boldface etc. replaced}
Defining the reflections for $\bm{w} = (w_1,w_2,w_3)$,
\begin{equation}    
\bm{\mathcal{R}}_2\bm{w} = (w_1,-w_2,w_3),\ \bm{\mathcal{R}}_3\bm{w} = (w_1,w_2,-w_3),
\end{equation}
the commutation property \eqref{commutation} gives
\begin{equation}    
[(\mathcal{L}_{(k,0,0)}+\mathbb{P}-\lambda)^{-1},\bm{\mathcal{R}}_j] = 0,\ j=1,2,
\end{equation}
and we conclude that the operator $(\mathcal{L}_{(k,0,0)}+\mathbb{P}-\lambda)^{-1}$ maps even/odd functions in $w_2$ or $w_3$ to even/odd functions in $w_2$ or $w_3$. This implies that any odd entries in $w_2$ or $w_3$ appearing in \eqref{GL} will vanish as they are integrated against the even Gaussian measure and $G$ has the structure
\begin{equation}\label{GLPast1}  
G= \tilde{{Q}}_{\bm{k}}
\begin{pmatrix}      
\ast & \ast & 0 & 0 & \ast\\        
\ast & \ast & 0 & 0 & \ast\\       
0 & 0 & \ast & 0 & 0\\        
0 & 0 & 0 & \ast  & 0\\        
\ast & \ast & 0 & 0 & \ast
\end{pmatrix}
\tilde{{Q}}_{\bm{k}}^T. 
\end{equation}

%\todo[inline]{Explanations below are not helpful: First, it refers to the H matrix, which is a separate section in the main text, and that is not even mentioned. Second, what is written just repeats what was cryptically already said in the main text. However, the derivation requires computing the Riesz projection as a contour integral, and explanation of that is totally missing. I think a reader of PRE will not be able to fill out this gap, so you need to help him.}
%This immediately implies that we can choose the basis vectors \eqref{basisshear} for the shear mode.
%\todo[inline, color = green]{Added an explanation why we do not actually evaluate the Riesz projection to the main text and moved proof of the structure of the first five entries of the eigenmodes to the corresponding section }

\section{Spectral temperature}
\label{app:Hmatrix}
Let us prove formula \eqref{basissimple} and \eqref{eq:spectemp} from the restricted Riesz projection \eqref{eq:Riesz} evaluated at the diffusion mode or the acoustic modes, i.e., the structure of the first five entries of a hydrodynamic eigenmode at wave number $k$.\\
First, we observe that, integrating $\mathcal{L}_{\bm{k}}f_\lambda=\lambda f_{\lambda}$ over $\bm{v}$ implies the relation
\begin{equation}\label{keta}  
-\ri \bm{k}\cdot(\eta_2,\eta_3,\eta_4) =  \lambda \eta_1,
\end{equation}
for ${\eta} = (\eta_1,\eta_2,\eta_3,\eta_4,\eta_5)$ being the corresponding column of $H$. We can normalize any eigenvector to have $\eta_1=1$ and, using \eqref{GLPast} in combination with \eqref{keta}, we arrive at
\begin{equation}    
[\tilde{Q}_{\bm{k}}{{\eta}_{\bm{k}}}(\lambda)]_2 = \frac{\ri}{k} \lambda,
\end{equation}
as well as $[\tilde{Q}_{\bm{k}}{{\eta}_{\bm{k}}}(\lambda)]_3=[\tilde{Q}_{\bm{k}}{{\eta}_{\bm{k}}}(\lambda)]_4=0$.\\

To evaluate $\eta_5$, we recall the definition of the adjugate matrix,  $M\text{adj}(M)=\det(M)I$, and we conclude that,
\begin{equation}
   { \text{adj}({G(\lambda,\bm{k})}-I)=c(\lambda, {\bm{k}})[\eta_{\bm{k}}(\lambda)\otimes\eta_{\bm{k}}(\lambda)],}
\end{equation}
up to a scalar multiplier {$c(\lambda,\bm{k})$} for $\lambda\in\{\lambda_{d},\lambda_{a},\lambda_{a}^*\}$, {since the eigenspace spanned by $\lambda$ is one-dimensional and the adjugate of $G(\lambda,\bm{k})-I$ has rank one}. Normalizing the first entry of $\eta$ to one as before implies that,
\begin{equation}
\eta_5(\lambda) =    {\theta(\lambda,\bm{k})} = \frac{\text{adj}[{G(\lambda,\bm{k})}-I]_{1,5}}{\text{adj}[{G(\lambda,\bm{k})}-I]_{1,1}},
\end{equation}
thus proving the general structure \eqref{basissimple} of the first five entries of an eigenmode associated to the diffusion or the acoustic modes.

From the definition of the spectral matrix \eqref{GLP} it immediately follows that,
\begin{equation}
    G(\lambda,\bm{k})^* =  G(\lambda^*,-\bm{k}).
\end{equation}
Consequently, using the linearity of the adjugate together with \eqref{GL},
\begin{equation}
\begin{split}
        \theta(\lambda,\bm{k})^* & = \left(\frac{\text{adj}[G(\lambda,\bm{k})-I]_{1,5}}{\text{adj}[G(\lambda,\bm{k})-I]_{1,1}}\right)^*\\
        & = \frac{\text{adj}[G(\lambda^*,-\bm{k})-I]_{1,5}}{\text{adj}[G(\lambda^*,-\bm{k})-I]_{1,1}}\\
        & = \frac{\text{adj}[G_{\rm red}(\lambda^*,k)-I]_{1,5}}{\text{adj}[G_{\rm red}(\lambda^*,k)-I]_{1,1}}\\
        & = \theta(\lambda^*,\bm{k}),
\end{split}
\end{equation}
where in the third step we have used that,
\begin{equation}
\begin{split}
    \text{adj}(G(\lambda^*,-\bm{k})-I) & =\text{adj}(\tilde{{Q}}_{\bm{k}}G_{\rm red}(\lambda^*,k)\tilde{{Q}}_{\bm{k}}^T-I)\\
     & =\text{adj}(G_{\rm red}(\lambda^*,k)-I).
\end{split}
\end{equation}

\section{Spectral properties of the linearized BGK model}
\label{app:BGK}
Here we summarize the pertinent findings of Refs.\ \cite{kogelbauerBGKspectral1,kogelbauerBGKspectral2}. For the linear BGK operator in frequency space,
\begin{equation}\label{defBGKfrequency}
\mathcal{L}_{\mathbf{k}}= -\ri \bm{v}\cdot \bm{k}-\frac{1}{\tau}+\frac{1}{\tau}\mathbb{P},
\end{equation}
the essential spectrum is the line $\{\Re\lambda = -\frac{1}{\tau}\}$ while
there exist exactly five eigenvalues (counted with multiplicity) above the essential spectrum for each wave number:
\begin{equation}\label{eq:spectrumBGK}
    \sigma_{\rm disc}(\mathcal{L}_{\bm{k}}) = \{\lambda_{\rm s}(k), \lambda_{\rm s}(k), \lambda_{\rm d}(k), \lambda_{\rm a}(k),\lambda_{\rm a}^*(k)\}.
\end{equation}
The degenerate shear mode, the real diffusion mode and the pair of complex conjugated acoustic modes \eqref{eq:spectrumBGK} are given as zeros of the spectral function,
\begin{equation}\label{spectralfunc}
    \Sigma(\lambda)=0.
\end{equation}
Spectral function $\Sigma$ is compactly written using the reduced wave number $\kappa=k\tau$ and $\zeta = \ri\frac{\tau\lambda+1}{\kappa}$ \cite{kogelbauerBGKspectral1},
%\begin{widetext}
%\begin{small}
    \begin{equation}\label{defSigma}
\begin{split}
&   \Sigma  = \frac{(Z(\zeta)-\ri\kappa)^2}{6(\ri \kappa)^5}
\left[
\zeta+6 \ri \kappa^3-\zeta  (\zeta^2+5) \kappa^2+2 \ri (\zeta ^2+3)\kappa \right.\\
&\left.-4 \ri Z^2 (\zeta )((\zeta ^2+1)\kappa -\ri \zeta)\right.\\
&\left. +Z(\zeta ) (\zeta ^2-(\zeta^4+4 \zeta ^2+11)\kappa^2+2 \ri  \kappa\zeta ^3 -5) ) \right],
%_{\zeta = \ri\frac{\tau\lambda+1}{\kappa}}.
\end{split}
\end{equation}
%\end{small}
%\end{widetext}
where
\begin{equation}\label{defZ}
    Z(\zeta) = \frac{1}{\sqrt{2\pi}} \int_{-\infty}^{\infty} \frac{e^{-\frac{v^2}{2}}}{v-\zeta }dv,
\end{equation}
is the plasma dispersion function \cite{fitzpatrick2014plasma}. 
The corresponding critical wave numbers for the branches \eqref{eq:spectrumBGK} are \cite{kogelbauerBGKspectral1},
\begin{equation}\label{kcrit}
    \begin{split}
       & \kappa_{\rm crit, s} = \sqrt{{\pi}/{2}}\approx 1.2533,\\
       &  \kappa_{\rm crit, a}  \approx 1.3118,\\
        &  \kappa_{\rm crit, d}\approx 1.3560.
    \end{split}
\end{equation}
Spectral temperature of the BGK equation is given explicitly by ($\zeta = \ri\frac{\tau\lambda+1}{\kappa}$) \cite{kogelbauerBGKspectral2},
%\begin{small}
\begin{equation}\label{eq:stBGK}
    \theta(\lambda)  = 
    %\left.
    \frac{\ri \sqrt{6} \left[\kappa^2 +\ri \zeta  \kappa+1 +Z(\zeta ) \left(\zeta +\ri \kappa \left(\zeta ^2+1\right)  \right)\right]}{\kappa \left(\zeta +\left(\zeta ^2-1\right) Z(\zeta )\right)}.
    %\right|_{\zeta = \ri\frac{\tau\lambda+1}{\kappa}}
\end{equation}
%\end{small}

%\todo[inline]{Here we can supply a picture of the transport coefficients for BGK, at least some information that "supports" the plots 2 and 3 in the main text} 
%\todo[inline,color = green]{Picture inserted with nicer style}

{Using algebraic-numerical solution for the eigenvalues from \eqref{spectralfunc} and \eqref{defSigma}, and also using the spectral temperature \eqref{eq:stBGK}, transport coefficients \eqref{transportcoef}, together with the shear mode, in dependence on wave number for the linear BGK equation \eqref{defBGKfrequency} are shown in Fig.\ \ref{prlplot}.

\begin{figure}
  \centering
   \includegraphics[width=1\linewidth]{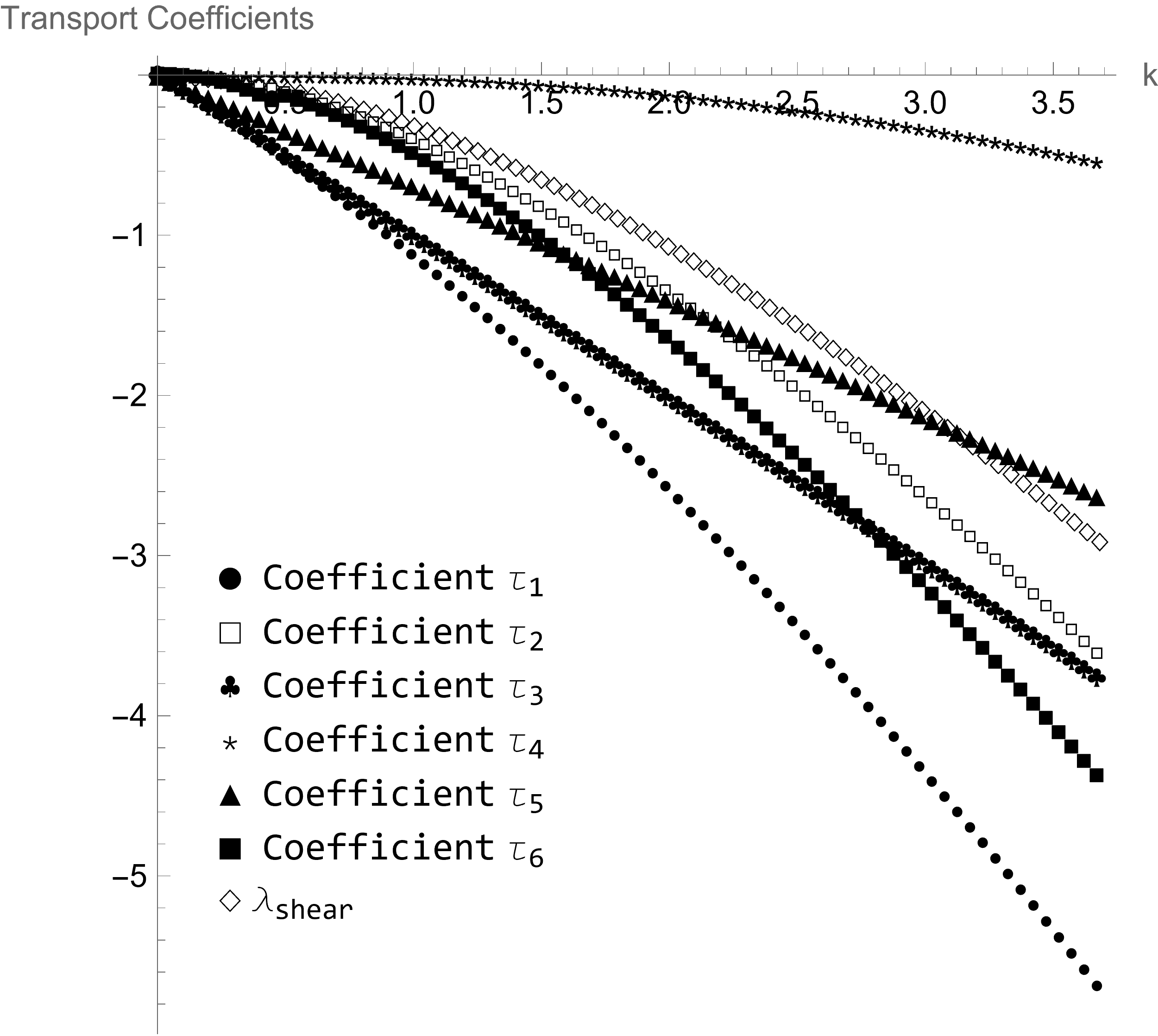}
   \caption{Transport coefficients \eqref{defT} evaluated for the BGK collision operator \cite{kogelbauerBGKspectral1} for Knudsen number $\tau=0.3$.  For small wave numbers, the generalized transport coefficients are approximated with those of the Navier--Stokes equation. }
    \label{prlplot}
\end{figure}

%The (minimal) critical wave number is associated to the shear mode and is given explicitly as
%\begin{equation}
% k_{\rm crit, shear} = \sqrt{\frac{\pi}{2}}\frac{1}{\tau}
%\end{equation}

\section{Gas flow in a channel}
\label{app:Kn}
For a uniform flow along the $z$-direction, we assume a constant pressure drop, which induces a constant forcing of the form 
\begin{equation}\label{forceuz}
   F_{\rm press}  = (0,0,0,f_{z,0},0), 
\end{equation}
for $f_{z,0}\sim\Delta p$, acting on the reduced hydrodynamics \eqref{dynh}:
\begin{equation}
 \partial_t \hat{{h}}= \mathcal{T} \hat{{h}} + F_{\rm press}.   
\end{equation}
Alternatively, one may regard \eqref{forceuz} as an external, uniform forcing along the flow direction. We refer to \cite{cercignani1963flow} for details of how a pressure drops translates to a constant forcing in the context of the 3D BGK equation.

Taking into account the no-slip boundary condition, we can expand \eqref{forceuz} using the Fourier series representation of a rectangular pulse,
\begin{equation}
  1|_{[-\frac{d}{2},\frac{d}{2}]}(x) = \frac{4}{\pi}\sum_{j=0}^{\infty} \frac{(-1)^j}{2j+1}\cos\left(\frac{2j+1}{2}\omega_x x\right),
\end{equation}
where $k=2j+1$ only ranges over odd wave numbers. We emphasize that - because of the Dirichlet boundary condition in the $x$-direction - the basic frequency for the Fourier elements in the $x$-direction is $\omega_x/2$. 

Taking averages in the $y$- and $z$-direction corresponds to setting $k_y=k_z=0$ for the frequencies in \eqref{defT} and \eqref{maineq} and we can expand the hydrodynamic variables as
\begin{equation}
    {h}(x,t) = (\rho_b,0,0,u_{z,0},T_b)+\sum_{j=1}^{\infty} \hat{{h}}_{j}\cos\left(\frac{2j+1}{2}\omega_x x\right),
\end{equation}
 where $k=|\bm{k}|=k_x=2j+1$ in the case of $y$-$z$-averaged dynamics, while $\rho_b$ and $T_b$ denote the equilibrium density and temperature at the wall. The constant base flow $u_{z,0}$ will be determined by requiring consistency to the Navier--Stokes equation with no-slip bounary condition (see subsequent calculation). 
 
 The rotation matrix,
 \begin{equation}%\label{defQ}
    \bm{Q}_{\bm{k}} =\left(
\begin{array}{ccc}
 \frac{k_x}{k} & -\frac{k_y}{k} & -\frac{k_z}{k} \\
 \frac{k_y}{k} & 1-\frac{k_y^2}{k^2+k_x k} & -\frac{k_y k_z}{k^2+k_x k} \\
 \frac{k_z}{k} & -\frac{k_y k_z}{k^2+k_x k} & 1-\frac{k_z^2}{k^2+k_x k} \\
\end{array}
\right),
\end{equation}
for the $y$-$z$-averaged dynamics ($k_y=k_z=0$) reduces to the identity matrix (same for $\bm{Q}_{\bm{\Omega}\bm{k}}$) and the equation for the steady state solution to \eqref{dynh} under the forcing \eqref{forceuz}  takes the form
\begin{equation}\label{steadyFourier}
{\mathcal{T}}\hat{{h}}^{\rm steady}_{j} +  (0,0,0,f_{z,0},0)\frac{4}{\pi}\frac{(-1)^j}{2j+1} = 0,\ k=2j+1.
\end{equation}
Equation \eqref{steadyFourier} can be solved readily to $\hat{{h}}_{j} = (0,0,0,\hat{q}_{j}^{\rm steady},0)$ to give the Fourier coefficients of the steady mass velocity,
\begin{equation}\label{soluz}
    \hat{q}_{j}^{\rm steady} = - \frac{4 f_{z,0}}{\pi\lambda_{\rm shear}[\omega_x/2(2j+1)]}\frac{(-1)^j}{2j+1}. 
\end{equation}
The steady flow rate is then given by
\begin{equation}\label{flowrate}
    F^{\rm steady} = - \frac{8 f_{z,0}}{\pi^2}\sum_{j=0}^\infty \frac{1}{\lambda_{\rm shear}[\omega_x/2(2j+1)](2j+1)^2},
\end{equation}
which, by the normalization \eqref{defQ} (taking into account that the pressure drop is related to the uniform force $f_{z,0}$) and the addition of the constant base flow, leads to \eqref{Qshear}.

To compare the dependence of the steady flow rate \eqref{flowrate} to the results in Cercignani \& Daneri \cite{cercignani1963flow}, we introduce the inverse Knudsen number as in \cite{cercignani1963flow},
%\todo[inline]{where do we see experimental results? Above, you call it inverse Knudsen number, why now some velocity parameter?
%Who is L?}
%\todo[inline, color=green]{Corrected}
\begin{equation}
\delta = \frac{d}{\tau} = \frac{2\pi}{\omega_x \tau}. 
\end{equation}
Since \eqref{flowrate} depends on frequency only through $k\tau$ and multiplication by $\omega_x$, the flow rate \eqref{defQ}, indeed, only depends on $\delta$.

Finally, to determine the base flow component $u_{z,0}$ in \eqref{Qshear}, we assume consistency with the steady, no-slip Navier--Stokes in the limit $\tau\to 0$ (vanishing relaxation time). Since for the Navier--Stokes solution for no-slip boundary condition is just the Poiseuille flow, the base-flow vanishes for $\tau\to 0$ and we find that, upon normalization of the basic frequency and the relaxation time,
\begin{equation}
\begin{split}
    u_{z,0} & = -\lim_{\tau\to 0} \frac{4\tau}{\pi^2}\sum_{j=0}^\infty \frac{1}{(2j+1)^2 \lambda_{\rm s}(2j+1)}\\
    & = -\lim_{\tau\to 0} \frac{4\tau}{\pi^2}\sum_{j=0}^\infty \frac{1}{(2j+1)^2[-\tau (2/\pi)^2 (2j+1)^2]}\\
    & =\sum_{j=0}^\infty \frac{1}{(2j+1)^4 },
    \end{split}
\end{equation}
where we have used the Navier--Stokes asymptotics of the BGK shear mode, 
\begin{equation}
    \lambda_{\rm s}(k)\sim -\tau k^2,
\end{equation}
evaluated at $k=\frac{2}{\pi}(2j+1)$, see \cite{kogelbauerBGKspectral2}.

%\label{commprop}

%\section{Explicit Formulas for the Transport Coefficients and Properties of the Spectral Closure}The transport coefficients appearing in \eqref{defT} are defined as expressions of the eigenvalues and spectral temperature evaluated on eigenvalues. We give the explicit dependencies for the determinant of the coordinate change from spectral to hydrodynamic variables,

\end{document}